\documentclass[a4paper,11pt]{amsart}
\usepackage{graphicx}
\usepackage{amssymb}
\begin{document}

\hyphenation{gra-vi-ta-tio-nal re-la-ti-vi-ty Gaus-sian
re-fe-ren-ce re-la-ti-ve gra-vi-ta-tion Schwarz-schild
ac-cor-dingly gra-vi-ta-tio-nal-ly re-la-ti-vi-stic pro-du-cing
de-ri-va-ti-ve ge-ne-ral ex-pli-citly des-cri-bed ma-the-ma-ti-cal
de-si-gnan-do-si coe-ren-za pro-blem gra-vi-ta-ting geo-de-sic
per-ga-mon cos-mo-lo-gi-cal gra-vity cor-res-pon-ding
de-fi-ni-tion phy-si-ka-li-schen ma-the-ma-ti-sches ge-ra-de
Sze-keres con-si-de-red tra-vel-ling ma-ni-fold re-fe-ren-ces
geo-me-tri-cal in-su-pe-rable sup-po-sedly at-tri-bu-table
Bild-raum in-fi-ni-tely counter-ba-lan-ces iso-tro-pi-cally
pseudo-Rieman-nian cha-rac-te-ristic geo-de-sics
Koordinaten-sy-stems ne-ces-sary Col-la-bo-ra-tion ine-qua-li-ties
coa-le-scence physi-cal ge-ne-ra-te}

\title[Remarks on numerical relativity, geodesic motions, \emph{etc.}] {{\bf Remarks on numerical relativity,\\geodesic motions,\\binary neutron star evolution}}

\author[Angelo Loinger]{Angelo Loinger}
\address{A.L. -- Dipartimento di Fisica, Universit\`a di Milano, Via
Celoria, 16 - 20133 Milano (Italy)}
\author[Tiziana Marsico]{Tiziana Marsico}
\address{T.M. -- Liceo Classico ``G. Berchet'', Via della Commenda, 26 - 20122 Milano (Italy)}
\email{angelo.loinger@mi.infn.it} \email{martiz64@libero.it}

\vskip0.50cm

\begin{abstract}
The computations of numerical relativity make use of $(3+1)-$
decompositions of Einstein field equations. We examine the
conceptual characteristics of this method; instances of
compact-star binaries are considered. The preeminent role of the
geodesic motions is emphasized.
\end{abstract}

\maketitle

\normalsize

\vskip1.20cm \noindent \textbf{1.} \emph{\textbf{Introduction.}}
In sect. \textbf{2} we recall properties and geodesics of the
Gauss\-ian-normal coordinate system. In sect. \textbf{3} we
emphasize that the physical results derived from any solution of
Einstein field equations must be independent of the adopted
reference frame. In sects. \textbf{4} and \textbf{5} we give a
r\'esum\'e of the properties of the $(3+1)$-decompositions of
Einstein equations, and we point out that: \emph{i}) the results
of the approximate computations obtained with the employment of
the gravitational energy-pseudotensor have an illusive value;
\emph{ii}) coordinate-system independence of the results is not
proved; \emph{iii}) no mathematical theorem of existence supports
the approximate computations of numerical relativity. Sect.
\textbf{5bis}: the particles of a \emph{discrete} ``cloud of
dust'' describe geodesic lines -- and therefore no GW is emitted
by them: an emblematic instance. Sects. \textbf{6} and \textbf{7}
contain some comments on the numerical computations concerning the
binaries composed of two ``Schwarschildian mass-points'' and of
two neutron stars; an analysis is given of the notions of
ADM-mass. Appendix A: The Einstein equations in a Gaussian-normal
frame. Appendix B: The equations of the standard
$(3+1)$-decomposition. Appendix C: The four-dimensional world as
``a mass of plasticine'' (Weyl), and some interesting
consequences.

\vskip1.20cm \noindent \textbf{2.} -- The concept of
Gaussian-normal coordinate system is due to Hilbert \cite{1}.
Landau and Lifshitz called it synchronous \cite{2}. Let us
consider in the four-dimensional world a three-dimensional space
$S_{3}$ such that every line in it is space-like; $(x^{1}, x^{2},
x^{3})$ be the point-coordinates in $S_{3}$. Starting from each
point $(x^{1}, x^{2}, x^{3})$, we trace the geodesic lines which
are orthogonal to $S_{3}$. They will become particular time-like
lines if we report on them the time $x^{0}=c\,t$ as the proper
time $\tau$. It follows easily that

\begin{equation} \label{eq:one}
\textrm{d}s^{2} = - (\textrm{d}x^{0})^{2} + g_{\alpha\beta} \,
\textrm{d}x^{\alpha} \textrm{d}x^{\beta} \, , \, (\alpha,
\beta=1,2,3) \quad.
\end{equation}

By virtue of the very construction of the three-dimensional space
$x^{0}=0$, the quadratic form $g_{\alpha\beta} \,
\textrm{d}x^{\alpha} \textrm{d}x^{\beta}$ is necessarily a
definite positive one.

\par It is interesting to write the differential equations of
\emph{all} the geodesic lines of the metric (\ref{eq:one}). The
Lagrangian $\mathcal{L}$:

\begin{equation} \label{eq:two}
\mathcal{L}:= -(\dot{x}^{0})^{2} + g_{\alpha\beta} \,
(x^{0},x^{1}, x^{2}, x^{3}) \, \dot{x}^{\alpha} \, \dot{x}^{\beta}
= Ac^{2} \quad,
\end{equation}

where the overdot denotes a derivative with respect to an affine
parameter $\sigma$, and $A$ is a constant, gives the Lagrangian
equations

\begin{equation} \label{eq:three}
\ddot{x}^{0} + \frac{1}{2} \, \frac{\partial
g_{\alpha\beta}}{\partial {x}^{0}} \,  \dot{x}^{\alpha} \,
\dot{x}^{\beta} = 0 \quad;
\end{equation}

\begin{equation} \label{eq:threeprime}
g_{\alpha\gamma} \, \ddot{x}^{\alpha} + \frac{\partial
g_{\alpha\gamma}}{\partial {x}^{\delta}} \, \dot{x}^{\alpha} \,
\dot{x}^{\delta} + \frac{\partial g_{\alpha\gamma}}{\partial
{x}^{0}} \, \dot{x}^{\alpha} \, \dot{x}^{0} - \frac{1}{2} \,
\frac{\partial g_{\alpha\beta}}{\partial {x}^{\gamma}} \,
\dot{x}^{\alpha} \, \dot{x}^{\beta} = 0 \quad, \quad (\gamma,
\delta =1,2,3) \,. \tag{3\'{}}
\end{equation}

$\mathcal{L}$ is a \emph{first integral} of
eqs.(\ref{eq:three}--\ref{eq:threeprime}); $A$ is negative,zero,
positive for time-like, null, space-like geodesics, respectively.
If $\sigma = \tau$, we have $A=-1$. The geodesics orthogonal to
$x^{0}=0$ are characterized by $0=\dot{x}^{\alpha} =
\ddot{x}^{\alpha}$, and $\sigma = \tau$.

\par The solutions of eqs.(\ref{eq:three}--\ref{eq:threeprime}) are commonly interpreted, if $\sigma = \tau$,
as the trajectories of test-particles. However, they represent
also the geodesic trajectories of the material elements of a
``cloud'', which generates the Gaussian field.

\par There are infinite expressions for the Gaussian-normal
$\textrm{d}s^{2}$, which are obtained by one of them with:
\emph{i}) any Lorentz transformation of the spacetime coordinates,
\emph{ii}) any transformation of the space coordinates.

\par Landau and Lifshitz \cite{2} write the Einstein field
equations in a Gaussian-normal frame; then, they separate space --
and time -- derivatives, thus obtaining a particular
($3+1)$--formalism. All the operations of raising and lowering of
the indices and the covariant derivatives are performed with
respect to the space metric $g_{\alpha\beta}
(x^{0},x^{1},x^{2},x^{3})$. They put $\kappa_{\alpha\beta}:=
\partial g_{\alpha\beta} / \partial t$, and in their equations
$(99,10)$, $(99,11)$, $(99,12)$ (see App. A) the symbol
$P_{\alpha}^{\beta}$ denotes the Ricci tensor with respect to the
\emph{three}-dimensional space. It is immediate to divide eqs.
$(99,10)$, $(99,11)$, $(99,12)$  in constraint equations,
containing $g_{\alpha\beta}$ and $\partial g_{\alpha\beta} /
\partial t$, and in evolution equations, containing $\partial^{2} g_{\alpha\beta} /
\partial t^{2}$. It can be proved \cite{3} that

\begin{equation} \label{eq:four}
\frac{\partial^{2} g_{\alpha\beta}}{\partial t^{2}} =
\frac{F_{\alpha\beta}}{g}
 \quad,
\end{equation}

where $g$ is the determinant of the Gaussian metric matrix, and
$F_{\alpha\beta}$ is an entire rational function of the
$g_{\alpha\beta}$'s, of the $\partial g_{\alpha\beta} /
\partial t$'s, of the second spatial and spatiotemporal
derivatives of $g_{\alpha\beta}$. (Of course, $F_{\alpha\beta}$
depends also on the matter tensor).

\par For the solution of the Cauchy problem it is necessary to
give the values of $g_{\alpha\beta}$ and $\partial g_{\alpha\beta}
/ \partial t$ at all the space points  $(x^{1},x^{2},x^{3})$ at an
initial instant $t=t_{0}$.

 \vskip1.20cm \noindent\textbf{3.} -- In principle, the Cauchy
 problem could be solved for \emph{any} physical system in \emph{any}
 system of coordinates, thus giving origin to an implicit,
 particular $(3+1)$--formulation of the problem.

 \par As it was emphasized by Hilbert \cite{1}, as a rule any
 solution of Einstein field equations must satisfy the following
 criterion: all the \emph{physical} results which we deduce from
 it must be \emph{independent} of the coordinate system in which
 the solution is expressed. Of course, this does not exclude the
 existence of interesting physical phenomena -- as, \emph{e.g.},
 the gravitational redshift of the spectral lines --, whose entity
 depends on the chosen reference system. An emblematic instance of
 independence of the coordinate system is represented by the
 \emph{geodesic} motions of the elements of a continuous ``cloud of
 dust''.

\vskip1.20cm \noindent\textbf{4.} -- The numerical relativity
makes use of two $(3+1)$-decompositions of the Einstein field
equations $[4$\emph{a}), \emph{b})$]$. The standard decomposition
is (see, \emph{e.g.}, p.$41$ of $[4$\emph{a})$]$):

\begin{equation} \label{eq:five}
\textrm{d}s^{2} = -\alpha^{2} \textrm{d}t^{2} + \gamma_{jk} \,
(\textrm{d}x^{j} + \beta^{j} \textrm{d}t) \, (\textrm{d}x^{k} +
\beta^{k} \textrm{d}t) \quad ,
\end{equation}

where $(j,k)=(1,2,3)$, and $c=G=1$. The matrix of the components
of metric (\ref{eq:five}) is:

\begin{displaymath} \label{eq:six}
{(-\alpha^{2}+\gamma_{jk}\beta^{j}\beta^{k}) \qquad
\gamma_{jk}\beta^{j} \choose \gamma_{jk}\beta^{k} \qquad \qquad
\qquad \gamma_{jk}} \quad. \tag{6}
\end{displaymath}

In Appendix B we report the field equations of the standard
decomposition.

\par If $\alpha=1$ and the three-vector $\beta^{j}$ is zero,
eq. (\ref{eq:five}) gives a Gaussian-normal $\textrm{d}s^{2}$. At
p.37 of $[4$\emph{a})$]$ we read: ``The freedom to choose these
four gauge functions $\alpha, \beta^{1}, \beta^{2}, \beta^{3}$
completely arbitrary embodies the fourfold coordinate degrees of
freedom inherent in general relativity ...'' The lapse function
$\alpha$ reflects the freedom to choose the sequence of the
space-like hypersurfaces, and the shift vector $\beta^{j}$
reflects the freedom to relabel the spatial coordinates on each of
the above hypersurfaces in an arbitrary way. This means that the
covariance of a $(3+1)$-decomposition is restricted to the
continuous transformations of of the space coordinates $x^{1},
x^{2},x^{3}$, and to the freedom to choose the initial
$t$-hypersurfaces. Consequently, \emph{the}
$(3+1)$-\emph{decompositions are} \textbf{\emph{not}}
\emph{equivalent to the Einstein field equations}, which possess
the general covariance with respect to \emph{all} the continuous
transformations of the spacetime coordinates $x^{0}, x^{1},
x^{2},x^{3}$. The $(3+1)$-decompositions are contrary to the
spirit of GR. And in them a vast class of coordinate systems is
not employable.

\vskip0.60cm \noindent\textbf{5.} -- In the applications to
various problems of the formalisms of the $(3+1)$--decompositions
of Einstein equations the authors choose suitable
particularizations of the metric (\ref{eq:five})--(\ref{eq:six}).
In these numerical computations an important role is commonly
played by the gravitational energy-pseudotensor, which is a tensor
only with respect to \emph{linear} coordinate transformations, and
therefore its intervention renders illusive the physical value of
any result. Indeed, it is not sufficient to remark that the total
pseudoenergy and the total pseudomomentum of the system form a
four-dimensional vector in the very distant Minkowskian spacetime.
Further, the authors omit to prove that the obtained results are
independent of the adopted reference system. No mathematical
theorem of existence is given: the numerical solutions presuppose
that ``suitable boundary conditions and initial data are chosen so
that these solutions do indeed exist'' ($[4$\emph{a})$]$, p.21).

\vskip0.60cm \noindent\textbf{5bis.} -- Articles and treatises of
numerical relativity and $(3+1)$-decomposi\-tions of the Einstein
equations contain many considerations and computations concerning
the properties and the generation of GWs. We have given several
demonstrations that GR, if properly understood, excludes the
\emph{physical} existence of GWs \cite{5} -- and the experience
corroborates our proofs. Therefore, we do not discuss here the
numerical-relativity methods regarding the GWs. We give instead
the simple proof that also the motions of the particles of a
\emph{discrete} ``cloud of dust'' are \emph{geodesic}. First of
all, according to a method by Infeld \cite{6} we write the
Einstein equations, in a generic coordinate system $x^{0},x^{1},
x^{2},x^{3}$, using tensor densities:

\setcounter{equation}{6}
\begin{equation} \label{eq:seven}
\sqrt{-g} \, R_{\mu\nu} \, - \frac{1}{2} \, g_{\mu\nu} \sqrt{-g}
\, R = -8\pi \sqrt{-g} \, T_{\mu\nu} \, ; \quad (\mu, \nu=0,1,2,3)
\, , \quad (c=G=1) \,.
\end{equation}

If $\delta(x-\xi):= \delta (x^{1}-\xi^{1}) \delta (x^{2}-\xi^{2})
\delta (x^{3}-\xi^{3})$ is the three-dimensional Dirac's
distribution, and $\xi^{1}, \xi^{2},\xi^{3}$ are the coordinate of
a particle, we put -- if the ``cloud'' is composed of $s$
particles:

\begin{equation}\label{eq:eigth}
\sqrt{-g} \,\,  T^{\mu\nu} = \sum_{p=1}^{s} \sqrt{-g} \, \,
\stackrel{p}{T}  {}^{\mu\nu} \quad,
\end{equation}

\begin{equation} \label{eq:nine}
\sqrt{-g} \,\, \stackrel{p}{T}  {}^{\mu\nu}  = \, \stackrel{p}{m}
(t) \, \,  \frac{\textrm{d} \stackrel{p}{\xi}
{}^{\mu}}{\textrm{d}s} \, \, \frac{\textrm{d} \stackrel{p}{\xi}
{}^{\nu}}{\textrm{d}s} \, \stackrel{p}{\delta} (x-\xi) \quad ;
\end{equation}

now, if the world-lines of the particles never intersect, it is
not difficult to verify that the equations of motion of the
particles:

\begin{equation} \label{eq:ten}
\sum_{p=1}^{s} ( \sqrt{-g} \, \, \stackrel{p}{T}
{}^{\mu\nu})_{;\nu} = 0 \quad,
\end{equation}

where the semicolon denotes a covariant derivative, are the
differential equations of geodesic lines. Consequently, no GW is
generated by our ``dust''; a result very general, because there
exists no limitation for the values of the particle
kinematical-elements (velocities, accelerations, time derivative
of the accelerations, \emph{etc}.).

\par We see that \emph{the gravitational self-force theory}, which
is based on a wrong analogy with the electromagnetic self-force of
a charge, does not make any physical sense. And \emph{the
post-Newtonian approximations} which concern a discrete ``cloud of
dust'' cannot give GWs.

\vskip0.60cm \noindent\textbf{6.} -- With regard to the binary
stars composed of two Schwarzschildian mass-points (see,
\emph{e.g.}, Chapt. 13 -- ``Binary Black Hole Evolution'' -- of
$[4$\emph{a})$]$), we observe that nobody has succeeded in proving
the theoretical existence of \emph{two} Schwarzschildian
mass-points, and therefore the \emph{approximate} solution of
Einstein equations obtained by numerical computations, which would
describe this system, has a very doubtful value. Further, we
remark that the black-hole interpretation of the Schwarzschildian
point-mass solution of Einstein equations has a mathematically
unfounded basis (see, \emph{e.g.}, sects. \textbf{3b}, \textbf{3c}
of \cite{5}). At any rate, the curvature (``hard'') singularity at
$r=0$, which characterizes the gravitational field of a point-mass
in a current consideration of the standard (Hilbert-Droste-Weyl)
form of the metric, can be simply removed if one adopts the
\emph{original} Schwarzschild's coordinate system, or Brillouin's
system, for which the metric has only a ``soft'' singularity at
$r=0$. Numerical stratagems, as the BH-excision, the moving
puncture method, \emph{etc}., are superfluous.

\vskip0.60cm \noindent\textbf{7.} -- The notion of ADM-mass (ADM:
Arnowitt-Descr-Misner) is currently employed in the numerical
computations $[4$\emph{a})$]$. Now, the definition of the ADM-mass
is founded on the properties of the gravitational
energy-pseudotensor (see sect. 5), and therefore the physical
meaning of an ADM-mass is rather uncertain.

\par Chapt. 16 of $[4$\emph{a})$]$ regards the ``Binary Neutron Star
Evolution''. The approximate computations make use of the above
mentioned notion of ADM-mass. We limit ourselves to emphasize that
the \emph{time evaluations} of the inspiral and merger phases of
the considered binaries and of the evolutions of the amplitudes
$h_{+}$ and $h_{\times}$ of hypothesized GWs (computed with the
disputable quadrupole formula) do not possess a physical value.
Indeed, these time intervals would make sense only if we could
prove their independence of the adopted coordinate system.

\vskip1.00cm
\par We conclude that the approximate computations of
numerical relativity are essentially self-referential, because no
experimental, or observational, proof and no theorem of
mathematical existence supports them. Further, they do not even
satisfy the criterion according to which the physical value of the
results depends on their independence of the reference frame.

\vskip2.00cm
\begin{center}
\noindent \small \emph{\textbf{APPENDIX A}}
\end{center}
\normalsize \noindent \vskip0.80cm

\par The Einstein field equations written in a Gaussian-normal
coordinate system (see sect. \textbf{2} and Landau and Lifshitz
\cite{2}) are $(\alpha, \beta =1,2,3)$:

\begin{equation} \label{eq:A1}
R_{0}^{0} = \frac{1}{2c} \, \frac{\partial
\kappa_{\alpha}^{\alpha}}{\partial t}  + \frac{1}{4} \,
\kappa_{\alpha}^{\beta} \kappa_{\beta}^{\alpha} = \frac{8\pi
G}{c^{4}} \, (T^{0}_{0} - \frac{1}{2} \, T) \quad, \tag{A1}
\end{equation}

\begin{equation} \label{eq:A2}
R_{\alpha}^{0} = \frac{1}{2} \, (\kappa_{\beta;\alpha}^{\beta} -
\kappa_{\alpha;\beta}^{\beta}) = \frac{8\pi G}{c^{4}} \,
T^{0}_{\alpha} \quad, \tag{A2}
\end{equation}

\begin{equation} \label{eq:A3}
R_{\alpha}^{\beta} = P_{\alpha}^{\beta} + \frac{1}{2c\sqrt{-g}} \,
\frac{\partial}{\partial t}  \, (\sqrt{-g} \,
\delta_{\alpha}^{\beta}) = \frac{8\pi G}{c^{4}} \,
(T^{\alpha}_{\beta} - \frac{1}{2} \, \delta^{\beta}_{\alpha} \, T)
\quad; \tag{A3}
\end{equation}

here: $\kappa_{\alpha\beta}: = \frac{\partial
g_{\alpha\beta}}{\partial t}$; $P_{\alpha}^{\beta}$ is the
three-dimensional Ricci tensor; the semicolon denotes a covariant
derivative with respect to the three-dimensional metric
$g_{\alpha\beta}$, $(\alpha, \beta =1,2,3)$.

\vskip2.00cm
\begin{center}
\noindent \small \emph{\textbf{APPENDIX B}}
\end{center}
\normalsize \noindent \vskip0.80cm

\par We give the basic formulae of the standard
$(3+1)$--decomposition of the Einstein field equations \cite{7}.

\par Index notations: $(i,j, \ldots)=(1,2,3)$; $(a,b,
\ldots=0,1,2,3)$. The extrinsic curvature $K_{ij}=\sqrt{-g^{00}}
\, \Gamma_{i}$$^{0}$$_{j}$ of a hypersurface measures how much
normal vectors to the hypersurface differ at neighboring points;
$K=\gamma_{ij} \, K^{ij}= K_{j}^{j}$. The symbol $D_{j}$ denotes
the covariant derivative with respect to $\gamma_{ij}$. Normal
vector: $n_{a}=(-\alpha,0,0,0)$; $n^{a}=(\alpha^{-1},
-\alpha^{-1}\beta^{1}, -\alpha^{-1}\beta^{2},
-\alpha^{-1}\beta^{3})$. The matter tensor $T_{ab}$ is the
fluidodynamical energy-tensor, with various polytropic equations
of state (EOSs); the matter source terms are:

\par $\varrho = n_{a} \, n_{b} \, T^{ab}; \quad
S^{i}=-\gamma^{ij}\, n^{a}\, T_{aj}; \quad S_{ij}= \gamma_{ia} \,
\gamma_{jb} \, T^{ab}; \quad S=\gamma^{ij} \, S_{ij} \quad.$

Metric $(c=G=1)$:

\begin{equation} \label{eq:B1}
\textrm{d}s^{2} = -\alpha^{2} \textrm{d}t^{2} + \gamma_{ij} \,
(\textrm{d}x^{i}+\beta^{i}\textrm{d}t) \,
(\textrm{d}x^{j}+\beta^{j}\textrm{d}t)\quad. \tag{B1}
\end{equation}

Two constraint equations ($R$ is the Ricci scalar):

\begin{equation} \label{eq:B2}
R + K^{2} - K_{ij} \, K^{ij} = 16\pi \rho \quad; \tag{B2}
\end{equation}

\begin{equation} \label{eq:B3}
D_{j} \, ( K^{ij}- \gamma^{ij}K) = 8\pi S^{i} \quad. \tag{B3}
\end{equation}

Evolution equation for the space metric $\gamma_{ij}$:

\begin{equation} \label{eq:B4}
\frac{\partial \gamma_{ij}}{\partial t}  = -2 \, \alpha \, K_{ij}
+ D_{i} \, \beta_{j} + D_{j} \, \beta_{i} \quad. \tag{B4}
\end{equation}

Evolution equation for the extrinsic curvature $K_{ij}$:

\begin{multline*} \label{eq:B5}
\frac{\partial K_{ij}}{\partial t} = \alpha \, (K_{ij} - 2 K_{ik}
\, K^{k}_{\phantom{i}j} + K K_{ij}) - D_{i} D_{j} \alpha - 8\pi
\alpha \left[ S_{ij} - \frac{1}{2}\, \gamma_{ij}(S-\varrho)
\right] +
\\ +\beta^{k} \, \frac{\partial K_{ij}}{\partial x^{k}} + K_{ik} \,
\frac{\partial \beta^{k}}{\partial x^{j}} +  K_{kj} \,
\frac{\partial \beta^{k}}{\partial x^{i}} \quad. \tag{B5}
\end{multline*}

\vskip2.00cm
\begin{center} \noindent \small \emph{\textbf{APPENDIX C}}
\end{center}
\normalsize \noindent \vskip0.80cm

\par We read in Weyl \cite{8} a penetrant analysis of a
fundamental property of the general-relativistic coordinate
systems. He wrote $[4$\emph{b})$]$: ``... the concept of relative
motion of several bodies has, as the postulate of general
relativity shows, no more foundation than the concept of absolute
motion of a single body. Let us imagine the four-dimensional world
as a mass of plasticine traversed by individual fibers, the world
lines of the material particles. Except for the condition that no
two world lines intersect, their pattern may be arbitrarily given.
The plasticine can then be continuously deformed so that not only
one but all fibers become vertical straight lines.'' It is clear
that this consideration implies that there exists always a
coordinate transformation, which allows us to pass from any
coordinate system for which some bodies are in motion to a
\emph{co-moving} coordinate system for which all these bodies are
at rest.

\par Now, no class of privileged coordinate systems exists in GR,
and any \emph{physical} effect must be frame independent \cite{1}.
Consequently, the fact that bodies at rest cannot generate
gravitational waves has a general significance: no coordinate
system exists for which the motions of the bodies generate
gravitational waves.

\par Remark that these considerations hold for the general case in
which both gravitational and non-gravitational forces are present.
For a different proof, founded on the Einstein field equations,
that \emph{all} the general-relativistic motions can be
\emph{geodesically} described, see our paper of ref. \cite{9}.

\par An immediate corollary: the gravitational field of a body,
whose motion is geodesic, moves \emph{en bloc} with the body, and
is propagated \emph{instantaneously}. In special relativity we
have  a partial analogue: the \emph{static}-electromagnetic fields
created by an electric charge in a rectilinear and uniform motion
(a Minkowskian geodesic motion) move \emph{along with} the charge,
and are propagated \emph{instantaneously} \cite{10}. (This
corresponds perfectly to the results found by an observer in a
rectilinear and uniform motion, who travels with respect to the
charge at rest).

\par This fact has been considered a paradox by some physicists,
who have tried to get rid of it with the gratuitous surmise that
it is hold only for infinite spatio-temporal motions of the
electric charge.

\par An instantaneous propagation of a field is not always in
contradiction with the theory of relativity. --

\end{document}